\documentclass[12pt]{article}

\usepackage[tbtags]{amsmath}
\usepackage{amssymb,eucal}

\DeclareMathOperator*{\eqls}{=}
\DeclareMathOperator{\tr}{tr}

\DeclareMathOperator{\arctg}{arctg}
\DeclareMathOperator{\tg}{tg}

\DeclareMathOperator{\cth}{cth}

\DeclareMathOperator{\ch}{ch}
\DeclareMathOperator{\sgn}{sgn}
\DeclareMathOperator{\fract}{fract}
\DeclareMathOperator{\integ}{integ}
\DeclareMathOperator*{\sint}{{\displaystyle\sum\mkern-25mu\int\,}}
\newcommand{\half}{\frac{1}{2}}
\newcommand{\J}[1]{{<\!\!\hat J\!\! >_{#1}}}

\title{Induced Angular Momentum in (2+1)-Dimensional Spinor 
Electrodynamics in Curved Space.
{\footnote{Published in \em Physics of Atomic Nuclei, Vol. 60, N 2, 1997, 
pp. 247-269.}}}
\author{Yu. A. Sitenko{\footnote{e-mail: yusitenko@gluk.apc.org}},
\qquad D. G. Rakityansky{\footnote{e-mail: radamir@ap3.gluk.apc.org}}\\
Bogolyubov Institute for Theoretical Physics,\\
National Academy of Sciences of Ukraine,\\
252143 Kiev, Ukraine}

\begin{document}

\maketitle

\begin{abstract}
Effects due to fermion-vacuum polarization by an external static magnetic 
field are considered in a two-dimensional noncompact curved space with a 
nontrivial topology. An expression for the vacuun angular momentum is 
obtained. Like the vacuum fermion number, it proves to be dependent on the 
global characteristics of the field and space.
\end{abstract}

\section*{\sc inrtoduction}
Vacuum quantum effects in strong external fields have long been the subject 
of intensive investigations (see, for example, \cite{Greeb,BirD}). It was 
found that fermion vacua can posess unusual properties and can be 
characterized by nontrivial quantum munbers. In particular, the fermion 
number can be nonzero and even fractional \cite{Jac76, Gold}. Fractional 
fermion numbers are realized in systems of various spatial dimensions. This 
effect may explain a number of phenomena, both in solid-state physics and in 
elementary-particle physics (for an overview, see \cite{NieS86}). Of special 
importance was the discovery of tthe fractional fermion number in 
(2+1)-dimensional spacetime, because it further resulted in the development 
of gauge-field models whose action functionals involve Chern-Simons 
topological term and in the formulation of some fruitful concepts such as 
anyons and fractional spin statistics, which made it possible to explain 
satisfactory some aspects of high-temperature superconductivity and 
fractional quantum Hall effect (see [6-9] and references cited in these 
studies).

In our opinion, however, analysis of the properties of the fermion vacuum has 
yet to completed. First, it is necessary to extend this analysis to quantum 
numbers other than the fermion one. Second, it is worth considering effects 
of the geometry of embedding space on these quantum numbers. The only 
attempts at advancements along these two lines were made in [10-12], where 
the vacuum angular momentum induced by an external magnetic field was found 
in a two-dimensional flat (Euclidean) space, and in \cite{Sit90,Sit91}, where 
the vacuum fermion number induced by an external magnetic field was 
determined in a two-dimensional curved (Riemann) space.

In this study, we find the vacuum angular momentum induced by an external 
axisymmetric magnetic field in a two-dimensional axisymmetric noncompact 
Riemann space with both trivial and nontrivial topologies.

\section*{\sc fermion facuum in static external fields}
        The two-dimensional Dirac Hamiltonian in the case considered has
the form
\begin{equation}\label{hamilton}
H=-i\alpha^\mu(x)[\partial_\mu + \frac{i}{2}\omega_\mu(x) - iV_\mu(x)]+\beta m,
\end{equation}
where
\begin{equation}\label{2}
\alpha^\mu(x)\alpha^\nu(x)=g^{\mu\nu}(x)+is\beta\varepsilon^{\mu\nu}(x),\quad
\mu,\nu=1,2,
\end{equation}
$g^{\mu\nu}$ and $\varepsilon^{\mu\nu}$ are the metric tensor and the totally
antisymmetric tensor of a surface, the values
$s=\pm 1$
mark two inequivalent irreducible representations of the Clifford algebra in
$2+1$-dimensional space-time,
$V_\mu$ is the
$U(1)$-bundle connection and $\omega_\mu$
is the spin connection.
The total magnetic flux (in the units of $2\pi$) through the surface
and the total integral curvature (in the units of $2\pi$) of the surface
can be presented in the form
\begin{gather}
\Phi=\frac{1}{2\pi}\int d^2\!x \sqrt{g}\varepsilon^{\mu\nu}\partial_\mu V_\nu
=\Phi^{(+)}-\Phi^{(-)},\notag\\   \label{3}
\Phi_K=\frac{1}{2\pi}s\beta\int d^2\!x \sqrt{g}\varepsilon^{\mu\nu}\partial_\mu \omega_\nu
=\Phi^{(+)}_K-\Phi^{(-)}_K,
\end{gather}
where
\begin{equation}\label{4}
\Phi^{(\pm)}=\lim_{\ln r\rightarrow\pm\infty}\frac{1}{2\pi}
\int\limits_0^{2\pi}d\theta V_\theta(r,\theta),\quad
\Phi^{(\pm)}_K=\lim_{\ln r\rightarrow\pm\infty}\frac{1}{2\pi}s\beta
\int\limits_0^{2\pi}d\theta \omega_\theta(r,\theta),
\end{equation}
$r$ and $\theta$ are the polar coordinates and $g=\det g_{\mu\nu}$; in eqs.
(\ref{3}) and (\ref{4}) we use the fact that a rotationally-symmetric
noncompact Riemann
surface has in general the topology of a cylinder.
Spinor wave function on such a surface is subject to the condition
\begin{equation}\label{psiamb}
\psi(r,\theta+2\pi)=e^{i2\pi\Upsilon}\psi(r,\theta).
\end{equation}
The Dirac Hamiltonian
$H$ (\ref{hamilton}) in the gauge
\begin{equation}\label{6}
V_r=0,\quad \partial_\theta V_\theta=0
\end{equation}
commutes with the operator
\begin{equation}\label{M}
M=-ix^\mu\varepsilon_\mu{}^\nu\partial_\nu-\Upsilon+\half s\beta.
\end{equation}
Hence a set of functions $\{\psi\}$ can be defined as the complete set
of solutions to the equations
\begin{equation}\label{8}
H\psi(x)=E\psi(x),\quad M\psi(x)=j\psi(x).
\end{equation}
If these functions satisfy the condition (\ref{psiamb}), then the eigenvalues
of the operator $M$ (\ref{M}) are half-integer,
\begin{equation}
j=n+\half s,\quad n\in \mathbb{Z},
\end{equation}
where $\mathbb{Z}$ is the set of integer numbers.  Thus eq. (\ref{M})
describes the angular momentum operator ---
the generator of rotations in spinor electrodynamics on a
rotationally-symmetric surface in a rotationally-symmetric external magnetic
field.

        Passing from the rotationally-symmetric gauge (\ref{6}) to
an arbitrary one, we get
\begin{equation}\label{10}
H\rightarrow e^{i\Lambda(x)}He^{-i\Lambda(x)},\quad
M\rightarrow e^{i\Lambda(x)}Me^{-i\Lambda(x)},
\end{equation}
where the gauge function $\Lambda(x)$ on a noncompact surface with the
topology of a cylinder satisfies the condition
\begin{equation}\label{11}
\Lambda(r,\theta+2\pi)=\Lambda(r,\theta)+2\pi\Upsilon_\Lambda.
\end{equation}
Taking into account
\begin{equation}\label{12}
V_\mu(x)\rightarrow V_\mu(x)+\partial_\mu\Lambda(x),\quad
\psi(x)\rightarrow e^{i\Lambda(x)}\psi(x),
\end{equation}
one can find
\begin{equation}\label{13}
\Phi^{(\pm)}\rightarrow \Phi^{(\pm)}+\Upsilon_\Lambda ,\qquad
\Upsilon\rightarrow\Upsilon+\Upsilon_\Lambda,
\end{equation}
while $\Phi$ and $\Phi^{(\pm)}-\Upsilon$ remain gauge invariant.

        Let us note that in a two-dimensional space, in distinction to a
three-dimensional one, the rotation group is abelian and the group-theoretical
arguments
restricting the eigenvalues of the angular momentum operator to half-integer
(and integer for bosons) numbers are lacking.  In the case of simply connected
surfaces (topologically equivalent to a plane) we have $\Upsilon=0$ and the condition
of single-valuedness of spinor wave functions, which is invariant under regular
($\Upsilon_\Lambda=0$) gauge transformations,  ensures that the eigenvalues
of the angular momentum operator are half-integer (and integer for bosons).
In the case of punctured surfaces (topologically equivalent to a cylinder),
at the same time with
eq. (\ref{M}), it is possible to define the angular momentum operator
alternatively (see, for example, \cite{1})
\begin{equation}\label{Malt}
M'=M-\Phi^{(-)}+\Upsilon,
\end{equation}
the eigenvalues of $M'$ (\ref{Malt}) on spinor functions satisfying eq.
(\ref{psiamb}) are not half-integer,
\begin{equation}
j'=n+\half s-\Phi^{(-)}+\Upsilon,\quad n\in \mathbb{Z}.
\end{equation}

The problem of choice between two possible definitions of the angular momentum
operator (eq. (\ref{M}) or (\ref{Malt})) is beyond the scope of current
investigation, since to solve this problem one has to take into account the
dynamics of the vector field $V_\mu$.
Let us only note here that in the case of a punctured plane
($\Phi_K^{(+)}=\Phi_K^{(-)}=0$) and the regular
(i.e. without the $\delta$-function singularity) part of the magnetic
field strength being equal to zero the option (\ref{M}) corresponds to the Maxwell
dynamics, and the option (\ref{Malt})  corresponds to the Chern-Simons
dynamics \cite{2,3}.

\section*{\sc angular momentum in (2+1)-dimensional\\ spinor electrodynamics}
        In the secondly quantized theory the operator of the dynamical
quantity corresponding to $M$ (\ref{M}) is defined in the conventional way
\begin{equation}\label{J}
\begin{split}
\hat J =&\half\int d^2x\sqrt{g}[\Psi^+(x),M\Psi(x)]_-=\\
=& \sint_E\sum_je^{-tE^2}
(a^+_{E,j}a_{E,j}-b^+_{E,j}b_{E,j})-\half\sint_E\sum_jje^{-tE^2}\sgn(E),
\end{split}
\end{equation}
where
\[
\sgn(u)=\left\{
\begin{array}{rl}
1,&u>0\\
-1&u<0
\end{array}\right.,
\]
$a^+_{E,j}$ and $a_{E,j}$ ($b^+_{E,j}$ and $b_{E,j}$) are the fermion
(antifermion) creation and annihilation operators satisfying the
anticommutation relations, the symbol $\sint\limits_E$ implies summation
over the
discrete and the integration (with a definite measure) over the continuous
parts of the energy spectrum and the regularization factor $\exp(-tE^2)$
($t>0$) is inserted to tame the divergence at $|E|\rightarrow\infty$.

        The operator of the dynamical quantity corresponding to $M'$
(\ref{Malt}) is defined as
\begin{equation}\label{Jalt}
\hat J'=\hat J-(\Phi^{(-)}-\Upsilon)\hat N,
\end{equation}
where $\hat N$ is the fermion number operator given by eq. (\ref{J}) with
the unity operator $\mathbb{I}$ substituted for $M$.

        The c-number piece of the angular momentum operator in the secondly
quantized theory J (\ref{J}) can be presented in the following way
\begin{equation}
\begin{split}
\J{}&=-\frac{1}{2\pi}\int\limits_{-\infty}^\infty du\sint_E\sum_j jE(E^2+u^2)^{-1}
e^{-tE^2}=\\
 &=-\frac{1}{2\pi}\int\limits_{-\infty}^{\infty}du\int dx\sqrt{g}\tr
<\!\! x|\frac{MHe^{-tH^2}}{H^2+u^2}|x\!\! >.
\end{split}
\end{equation}

Using the relation
\[
[M,\beta]_-=0
\]
and the trace identity
\begin{multline}
\frac{i}{2}g^{-\half}\partial_\mu g^{\half}\tr<\!\! x|\alpha^\mu
\frac{M\beta H_0}{H_0^2+m^2+u^2}e^{-tH_0^2}|x\!\! >=
\\=\tr<\!\! x|M\beta e^{-tH_0^2}|x\!\! >-(m^2+u^2)\tr<\!\! x|
\frac{M\beta e^{-tH_0^2}}{H_0^2+m^2+u^2}|x\!\! >,
\end{multline}
where $H_0=H|_{m=0}$, we get
\begin{multline}
\J{}=-\half\sgn(m)e^{-t m^2}\int d^2x\sqrt{g}\tr<\!\! x|\beta M
e^{-t H_0^2}|x\!\! >+\\
+\frac{im}{4\pi}e^{-tm^2}\int\limits_{-\infty}^{\infty}\frac{du}{m^2+u^2}
\int\limits_\sigma\tr<\!\! x|\frac{M\beta H_0 e^{-tH_0^2}}{H_0^2+m^2+u^2}|x\!\! >
\varepsilon_{\mu\nu}(x)dx^\nu,
\end{multline}
where $\sigma$ is the closed contour conditionally bounding a noncompact
two-dimensional space (surface) at infinity.  We recall that for the surfaces
with nontrivial topology the contour $\sigma$ can consist of several disjoint
components: for example, if for the infinite plane (trivial
topology) the contour $\sigma$ is a circle of infinite radius (one-component
boundary at
infinity), then for the cylinder of infinite length and finite radius
(non-trivial topology) the contour $\sigma$ consists of two circles of finite radius
(two-component boundary at infinity).

        In the case of a noncompact surface with the topology of a cylinder
we get
\begin{equation}
\J{}=-\half e^{-tm^2}\left[\sgn(m)A^{(M)}(t)+S^{(M)}_+(m,t)-S^{(M)}_-(m,t)
\right],
\end{equation}
where
\begin{equation}
A^{(M)}(t)=\int d^2x \sqrt{g}\tr<\!\! x|\beta Me^{-tH_0^2}|x\!\! >
\end{equation}
and
\begin{multline}
S^{(M)}_\pm(m,t)=-\frac{sm}{2\pi}\int\limits_{-\infty}^\infty
\frac{du}{m^2+u^2}\int\limits_0^{2\pi} d\theta\\
\tr<\!\! \ln r\rightarrow\pm\infty,\theta|r\sqrt{g}\alpha^\theta
\frac{H_0Me^{-tH_0^2}}{H_0^2+m^2+u^2}|\ln r\rightarrow\pm\infty,\theta\!\! >.
\end{multline}
We prove the existence of the asymptotical expansion
\[
\J{}\eqls_{t\rightarrow 0_+}t^{-1}\sum_{l=0}^\infty C_lt^{\frac{l}{2}},
\]
and calculate the coefficients corresponding to $l=0,1,2$ (details will be
published elsewhere).  It turns out that the coefficients $C_0$ and $C_1$
are depending on $\Phi_K^{(\pm)}$ but independent of $\Phi^{(\pm)}$ and
$\Upsilon$.  The latter circumstance allows us to
define the renormalized vacuum value as
\begin{equation}
\J{\rm ren}=
\lim_{t\rightarrow 0_+}\left(\J{}-\J{}\mid_{\Phi^{(\pm)}=\Upsilon=0}\right).
\end{equation}
It is clear that this definition ensures the normal ordering of the operator
product in the non-interacting theory
($\J{\rm ren}|_{\Phi^{(\pm)}=\Upsilon=0}=0$).

        Let us present the final form for the renormalized vacuum value of angular mometum
on a surface with the topology of a cylinder:
\begin{multline}\label{113}
\J{\rm ren}=-\frac{1}{4}s\sgn(m)\left[\left(\Phi^{(+)}-\Upsilon\right)^2-
\left(\Phi^{(-)}-\Upsilon\right)^2\right]
-\\-
\half\left(\Phi^{(+)}-\Upsilon\right)
\xi_+\left[m,\fract\left(\Phi^{(+)}-\Upsilon+\half\right),\Phi_K^{(+)}\right]
-\\-
\frac{1}{4}\tau_+\left[m,\fract\left(\Phi^{(+)}-\Upsilon+\half\right),\Phi_K^{(+)}\right]
+\\+
\half\left(\Phi^{(-)}-\Upsilon\right)
\xi_-\left[m,\fract\left(\Phi^{(-)}-\Upsilon+\half\right),\Phi_K^{(-)}\right]
+\\+
\frac{1}{4}\tau_-\left[m,\fract\left(\Phi^{(-)}-\Upsilon+\half\right),\Phi_K^{(-)}\right],
\end{multline}
where
\begin{equation}
\xi_\pm\left(m,v,\Phi^{(\pm)}_K\right)=\left\{
\begin{array}{lr}
0,&\Phi^{(\pm)}_K\lessgtr 1\\
s\left\{\arctg\left[\cth(\pi R_\pm m)\tg(\pi v)\right]-\sgn(m)v\right\},&\Phi^{(\pm)}_K=1\\
s\sgn(m)\left[\half\sgn_0(v)-v\right],&\Phi^{(\pm)}_K\gtrless 1
\end{array}%
\right.,%
\end{equation}%
\begin{equation}%
\tau_\pm\left(m,v,\Phi^{(\pm)}_K\right)=\left\{%
\begin{array}{lr}%
0,&\Phi^{(\pm)}_K\lessgtr 1\\
s\left\{-
\frac{2}{\pi}\int\limits^v_{1/2}
du\,\arctg\left[\cth(\pi R_\pm m)\tg(\pi u)\right]+\right.\\
+\sgn(m)\left(v^2-\frac{1}{4}\right)+
\left.\vphantom{\int\limits_{1/2}^v}\frac{1}{\pi}R_\pm m\ln\left[
1-\frac{\cos^2(\pi v)}{\ch^2(\pi R_\pm m)}\right]\right\},&\Phi^{(\pm)}_K=1\\
s\sgn(m)\left[\half\sgn(v)-v\right]^2,&\Phi^{(\pm)}_K\gtrless 1
\end{array}
\right.
\end{equation}
and we introduce the following notations
\[
\fract(u)=u-\integ(u),\quad\sgn_0(u)=\left\{
\begin{array}{ll}
\sgn(u),&u\ne0\\
0,&u=0
\end{array}\right.,
\]
${\rm integ}(u)$ is the nearest to  $u$ integer;
\[
-\half<\fract(u)<\half,\quad \lim_{\varepsilon\rightarrow 0_+}
\integ(n+\half\pm\varepsilon)=n+\half\pm\half.
\]
Note that the functions $\xi_+$ and $\xi_-$ which determine the boundary
contribution to the vacuum value of the fermion number on a surface with the
topology of a cylinder have been found earlier in \cite{4} (where they are
denoted as $\xi$ and $\xi_0$ respectively).
In the case of the alternative definition of angular momentum (\ref{Jalt})
we get
\begin{multline}\label{114}
<\!\!\hat J'\!\! >_{\rm ren}\equiv\J{\rm ren}
-\left(\Phi^{(-)}-\Upsilon\right)<\!\!\hat N\!\! >_{\rm ren}
=-\frac{1}{4}s\sgn(m)\Phi^2
-\\\shoveright{-
\half\Phi\xi_+\left[m,\fract\left(\Phi^{(+)}-\Upsilon+\half\right),\Phi_K^{(+)}\right]
-\\-
\frac{1}{4}\tau_+\left[m,\fract\left(\Phi^{(+)}-\Upsilon+\half\right),\Phi_K^{(+)}\right]
+\hphantom{,}}\\+
\frac{1}{4}\tau_-\left[m,\fract\left(\Phi^{(-)}-\Upsilon+\half\right),\Phi_K^{(-)}\right].\hphantom{+}
\end{multline}
The latter expression at fixed values of $\Phi$ depends on
$\Phi^{(-)}-\Upsilon$ ( or $\Phi^{(+)}-\Upsilon$) periodically with
the period equal to 1, which is also peculiar for the renormalized vacuum value of
fermion number on a surface with the topology of a cylinder \cite{Sit90,Sit91}
\begin{equation}\label{115}
\begin{split}
<\!\!\hat N\!\! >_{\rm ren}\equiv&\lim_{t\rightarrow 0_+}<\!\!\hat N\!\! >=\\
=&-\frac{1}{2}s\sgn(m)\Phi
-\half\xi_+\left[m,\fract\left(\Phi^{(+)}-\Upsilon+\half\right),\Phi_K^{(+)}\right]
+\\\hphantom{-\frac{1}{2}s\sgn(m)\Phi}&+
\half\xi_-\left[m,\fract\left(\Phi^{(-)}-\Upsilon+\half\right),\Phi_K^{(-)}\right].
\end{split}
\end{equation}

Let us emphasize that the results obtained are invariant under gauge
transformations, including singular ones,
$\Upsilon_\Lambda\neq 0$ (\ref{11}).
The gauge invariance of (\ref{113}) and (\ref{114})
is stipulated by the choice of the operators  $M$ (\ref{M})  and
$M'$ (\ref{Malt})
in the capacity of the angular momentum operator in the first-quantized theory:
although the operators $M$  ¨ $M'$ are changed
(covariantly, as well as the Hamiltonian $H$ (\ref{hamilton}) does, see eq.(\ref{10}))
under gauge transformations,
their eigenvalues, as well as the values of the energy, remain gauge invariant.
Expressions (\ref{113}), (\ref{114}) and (\ref{115}) are also invariant under simultaneous
substitution $s\rightarrow -s$ and $m\rightarrow -m$, which means
that the transition to inequivalent representation can be implemented
by changing the sign of the fermion mass.

In the case of a simply connected surface  we have
\begin{align}
\Phi^{(-)}&=0,&\Phi^{(-)}_K&=0,&\Upsilon&=0,\notag\\
\Phi^{(+)}&=\Phi,&\Phi^{(+)}_K&=\Phi_K,&&
\end{align}
and the expressions for the renormalized vacuum values take the form
\begin{align}
<\!\!\hat N\!\! >_{\rm ren}=&-\frac{1}{2}s\sgn(m)\Phi
-\half\xi_+\left[m,\fract\left(\Phi+\half\right),\Phi_K\right]\\
\intertext{and}
\J{\rm ren}=&-\frac{1}{4}s\sgn(m)\Phi^2-
\half\Phi\xi_+\left[m,\fract\left(\Phi+\half\right),\Phi_K\right]-\notag\\
&-\frac{1}{4}\tau_+\left[m,\fract\left(\Phi+\half\right),\Phi_K\right].
\end{align}
It is instructive to compare the results in two very simple
particular cases differing in topology.
In the case of a plane we have the vacuum fermion number \cite{6}
\begin{align}
<\!\!\hat N\!\! >_{\rm ren}=&
-\frac{1}{2}s\sgn(m)\Phi\\
\intertext{and the vacuum angular momentum [13-15]}
\J{\rm ren}=&
-\frac{1}{4}s\sgn(m)\Phi^2.
\end{align}
In the case of a punctured plane and $\Phi=0$ we have the vacuum fermion
number \cite{Sit90,Sit91}
\begin{multline}
<\!\!\hat N\!\! >_{\rm ren}=
\frac{1}{2}s\sgn(m)\left\{\half\sgn_0\left[\fract\left(\Phi^{(-)}-\Upsilon+\half
\right)\right]-\right.\\
\left.-\fract\left(\Phi^{(-)}-\Upsilon+\half\right)\right\}
\end{multline}
and the vacuum angular momentum
\begin{multline}
\J{\rm ren}=
\frac{1}{2}s\sgn(m)\left(\Phi^{(-)}-\Upsilon\right)
\left\{\half\sgn_0\left[\fract\left(\Phi^{(-)}-\Upsilon+\half
\right)\right]-\right.\\
\left.-\fract\left(\Phi^{(-)}-\Upsilon+\half\right)\right\}+\\
+\frac{1}{4}s\sgn(m)\left\{\half\sgn\left[\fract\left(\Phi^{(-)}-\Upsilon+\half
\right)\right]-\right.\\
\left.-\fract\left(\Phi^{(-)}-\Upsilon+\half\right)\right\}^2
\end{multline}
or
\begin{multline}
<\!\!\hat J'\!\! >_{\rm ren}=
\frac{1}{4}s\sgn(m)\left\{\half\sgn\left[\fract\left(\Phi^{(-)}-\Upsilon+\half
\right)\right]-\right.\\
-\left.\fract\left(\Phi^{(-)}-\Upsilon+\half\right)\right\}^2.
\end{multline}

In conclusion we note that the vacuum fermion number
induced on a noncompact Riemann surface of a more general form --- topologically
finite orientable surface (possessing a disconnected multi-component boundary
at infinity and handles) --- was calculated in \cite{Sit90,Sit91}.
The inducing of the vacuum angular momentum on such a surface will be
considered elsewhere.

        We are thankful to P. I. Fomin and V. P. Gusynin for stimulating
discussions.  The research was in part supported by the International Science
Foundation (grant GNZ 000).

\end{document}